\documentclass[11pt]{amsart}

\usepackage{amsxtra}

\newtheorem{theorem}{Theorem}[section]
\newtheorem{proposition}[theorem]{Proposition}
\newtheorem{lemma}[theorem]{Lemma}
\newtheorem{corollary}[theorem]{Corollary}
\theoremstyle{definition}

\theoremstyle{remark}
\newtheorem{remark}[theorem]{Remark}

\numberwithin{equation}{section}

\newcommand{\N}{\mathbb{N}}
\newcommand{\Z}{\mathbb{Z}}

\newcommand{\R}{\mathbb{R}}
\newcommand{\C}{\mathbb{C}}

\DeclareMathOperator{\curl}{curl}
\DeclareMathOperator{\dist}{dist}
\DeclareMathOperator{\ran}{ran}
\DeclareMathOperator{\supp}{supp}

\begin{document}

\title[On the asymptotic number of edge states]{On the asymptotic
  number of edge states for magnetic Schr\"odinger operators}

\author{Rupert L. Frank}
\address{Royal Institute of Technology, Department of Mathematics, 100
  44 Stockholm, Sweden}
\email{rupert@math.kth.se}
\date{\today}

\begin{abstract}
  We consider a Schr\"odinger operator $(h\mathbf D -\mathbf A)^2$
  with a positive magnetic field $B=\operatorname{curl}\mathbf A$ in
  a domain $\Omega\subset\R^2$. The imposing of Neumann boundary
  conditions leads to spectrum below $h\inf B$. This is a boundary
  effect and it is related to the existence of edge states of the
  system.

  We show that the number of these eigenvalues, in the semi-classical
  limit $h\to 0$, is governed by a Weyl-type law and that it involves
  a symbol on $\partial\Omega$. In the particular case of a constant
  magnetic field, the curvature plays a major role.
\end{abstract}

\maketitle

%%%%%%%%%%%%%%%%%%%%%%%%%%%%%%%%%%%%%%%%%%%%%%%%%%%%%%%%%%%%%%%%%%%%%%%%%%%%%%%
%%%%%%%%%%%%%%%%%%%%%%%%%%%%%%%%%%%%%%%%%%%%%%%%%%%%%%%%%%%%%%%%%%%%%%%%%%%%%%%

\section{Introduction and main results}

\subsection{Introduction}

In this paper we consider a magnetic Schr\"odinger operator
\begin{equation}\label{eq:op}
  P_h := (h\mathbf{D}-\mathbf{A})^2
  \qquad \text{in } L_2(\Omega)
\end{equation}
where $\mathbf D=-i\nabla$ and $\Omega\subset\R^2$ has a smooth and
compact boundary $\partial\Omega$ on which we impose Neumann boundary
conditions. The 'magnetic field' $\curl\mathbf{A}=B$ is assumed to be
smooth, positive and 'not too large on the boundary' (see \eqref{eq:b}
below for the precise assumption). As we will see, the choice of
Neumann boundary conditions implies the existence of eigenvalues below
$h\inf B$. Our goal is to determine for $b_0<\inf B$ the asymptotic
behavior of the number
\begin{equation}\label{eq:number}
  N(hb_0,P_h)
\end{equation}
of eigenvalues of $P_h$ below $hb_0$ in the semi-classical limit
$h\rightarrow 0$. The semi-classical limit is of course equivalent to
the limit of a strong magnetic field.

The operator \eqref{eq:op} has recently received a lot of attention in
connection with the Ginzburg-Landau theory of
superconductivity. Indeed, the lowest eigenvalue of \eqref{eq:op}
plays an important role in the description of the boundary nucleation
of superconductity close to the third critical field, see \cite{LuPa},
\cite{HeMo}, \cite{FoHe2} and references therein. The methods developed
in these papers will also be important for our analysis. They show in
particular that the boundary has an effect similar to that of a
potential well and that eigenfunctions of $P_h$ corresponding to
eigenvalues below $h\inf B$ are strongly localized near the
boundary. Hence they correspond to 'edge states'.

We emphasize that the 'energy level' in \eqref{eq:number} depends on
$h$. If instead the energy $\lambda>0$ is fixed, it is well-known that
\begin{equation*}
  N(\lambda,P_h) \sim h^{-2} \lambda \frac{|\Omega|}{4\pi}
\end{equation*}
for bounded $\Omega$. In particular, the leading term is independent
of the magnetic field. Recalling that the distance between two Landau
levels is proportional to $h$ it is natural to expect effects of the
magnetic field to appear when $\lambda=b_0 h$. We note also that
\eqref{eq:number} can be rewritten as
\begin{equation*}
  N(hb_0,P_h) = N(b_0,(h^{1/2}\mathbf{D}-h^{-1/2}\mathbf{A})^2).
\end{equation*}
This suggests that \emph{the effective semi-classical parameter of the
problem is $h^{1/2}$}, which tends to zero simultaneously as the
effective magnetic field $h^{-1/2} B$ tends to infinity.

The first results on spectral asymptotics of the operator
\eqref{eq:op} were obtained by Colin de Verdi\`ere \cite{CV} and
Tamura \cite{Ta}. The former author considers a slightly different
problem, namely the behavior of $N(\lambda,P_1)$ as
$\lambda\to\infty$. Moreover, $P_1$ is defined there as the
\emph{Dirichlet} realization of \eqref{eq:op}. The methods of
\cite{CV} (see also \cite{Tr}) allow to prove that under suitable
assumptions on $\mathbf A$ one has for all $b_0>0$
\begin{equation}\label{eq:cv}
  N(hb_0,P_h) \sim 
  h^{-1}\frac1{2\pi}\sum_{n=1}^\infty 
  \int_{\{x\in\Omega: \ (2n-1)B(x)<b_0\}} B(x)\, dx.
\end{equation}
This formula is valid both for the Dirichlet and for the Neumann
realization. However, while the Dirichlet realization has no spectrum
below $h\inf B$, this is no longer true for the Neumann
realization. In this case \eqref{eq:cv} provides the estimate
\begin{equation*}
  N(hb_0,P_h) = o(h^{-1}), \qquad b_0\leq \inf B.
\end{equation*}
Our goal is to improve upon this and to determine the precise behavior
of $N(hb_0,P_h)$ as $h\to 0$. So loosely speaking we are interested in
the leading term of the spectral asymptotics whose first term
vanishes.

For further results on spectral asymptotics of magnetic Schr\"odinger
operators we refer to \cite{LiSoYn}, \cite{Iv}, \cite{DiRa} and
references therein.

%%%%%%%%%%%%%%%%%%%%%%%%%%%%%%%%%%%%%%%%%%%%%%%%%%%%%%%%%%%%%%%%%%%%%%%%%%%%%%%

\subsection{Main result}\label{sec:main}

Let us state the precise assumptions on $\Omega$ and $\mathbf A$ under
which we shall work henceforth. We assume that $\Omega\subset\R^2$ is
an open domain with $\partial\Omega\in C^4$. Moreover, for the sake of
simplicity, we assume that $\partial\Omega$ is bounded and
connected. Note that we do \emph{not} assume that $\Omega$ is bounded
or simply connected. We consider a `magnetic vector potential'
$\mathbf A\in C^2(\overline\Omega,\R^2)$ and introduce the `magnetic
field' $B:=\curl\mathbf A$ and the quantities
\begin{equation*}
  b:= \inf_{x\in\Omega}B(x), \qquad b':= \inf_{x\in\partial\Omega}B(x).
\end{equation*}
The operator \eqref{eq:op} is defined via the quadratic form
\begin{equation*}
  q_h[u]:=\int_\Omega |(h\mathbf D -\mathbf A)u|^2\,dx  
\end{equation*}
with domain
\begin{equation*}
  \mathcal D[q_h] := \{u\in L_2(\Omega) : \ 
  (h\mathbf D -\mathbf A)u \in L_2(\Omega,\C^2) \}.
\end{equation*}

If the essential spectrum of $P_h$ is non-empty, which may happen if
$\Omega$ is unbounded, we use the well-known inequality
\begin{equation}\label{eq:dirichletineq}
  hb\int_\Omega |u|^2\,dx \leq 
  \int_\Omega |(h\mathbf D -\mathbf A)u|^2\,dx,
  \qquad u\in C_0^\infty(\Omega),
\end{equation}
and a `magnetic version' of Persson's lemma to conclude that
$\inf\sigma_{ess}(P_h)\geq hb$. Hence in any case for $\lambda <hb$
the spectrum of $P_h$ in the interval $[0,\lambda)$ consists of
finitely many eigenvalues of finite multiplicities, and we denote
their total number (taking multiplicities into account) by $N(\lambda,P_h)$. 
  
In order to state our main results we need some notation. For
$\xi\in\R$ we denote by $\mu(\xi)$ the lowest eigenvalue of the
operator
\begin{equation*}\label{eq:modelopintro}
  -\frac{d^2}{dt^2}+(\xi+t)^2 \qquad\text{in } L_2(\R_+)
\end{equation*}
with a Neumann boundary condition at the origin. Then (see Subsection ~
\ref{sec:ode} for more details) the minimum
\begin{equation*}
  \Theta_0 := \inf_{\xi\in\R}\mu(\xi)
\end{equation*}
is attained at a unique $\xi_0$ and one has $\xi_0\in (-1,0)$ and 
\begin{equation}\label{eq:c1}
  \mathcal C_1 := \mu''(\xi_0)/6 |\xi_0|>0.
\end{equation}

Throughout the following we shall assume that
\begin{equation}\label{eq:b}
  0<\Theta_0b'<b.
\end{equation}
Since $0<\Theta_0<1$ (numerically $\Theta_0=0.59\dots$) this is in
particular true in the important special case of a constant magnetic
field.

Our first main result is the following. 

\begin{theorem}\label{main1}
  Let $\Theta_0b'<b_0<b$. Then
  \begin{equation}\label{eq:main}
    \lim_{h\to0} h^{1/2} N(hb_0,P_h) =
    \frac{1}{2\pi} 
    \iint_{\{(x,\xi)\in\partial\Omega\times\R:\ B(x)\mu(\xi)<b_0\}}
    B(x)^{1/2}\, ds(x)d\xi. 
  \end{equation}
\end{theorem}

We emphasize that \eqref{eq:main} has a Weyl-type form, involving the
symbol $B(x)\mu(\xi)$ on the co-tangent bundle of the one-dimensional
manifold $\partial\Omega$. The essentially \emph{one}-dimensional
nature of the asymptotics is also reflected in the fact that the
effective semi-classical parameter $h^{1/2}$ appears with the power
$-1$ in the asymptotics. This should be compared with \eqref{eq:cv}
for $b_0> b$. There both the integral term and the power of $h^{1/2}$
reflect the \emph{two}-dimensional nature of the bulk states.

Note that Theorem \ref{main1} implies that
$N(h\Theta_0b',P_h)=o(h^{-1/2})$. It is natural to ask whether it is
possible to obtain the correct asymptotics. This will probably involve
the geometry of the set $\{x\in\partial\Omega:\ B(x)=b'\}$. Here we
give an answer in the particular case where the magnetic field $B$ is
constant. Indeed, we do not only give the asymptotics of $N(hb_0,P_h)$
for $b_0=\Theta_0 B$ but we allow $b_0$ to vary with $h$ on the scale
$h^{1/2}$. The result will involve the curvature
$\kappa:\R/|\partial\Omega|\Z\to\R$, see Subsection
\ref{sec:bdrycoord} for our notation. We follow the usual convention
that $\kappa\geq 0$ if $\Omega$ is convex. We will prove

\begin{theorem}\label{main2}
  Assume that $B$ is constant and let $\kappa_0\in\R$. Then
  \begin{equation}\label{eq:main2}
    \lim_{h\to0} h^{1/4} N(h\Theta_0B+
    h^{3/2}\mathcal C_1B^{1/2}\kappa_0,P_h)
    = \frac{B^{1/4}}{\pi\sqrt{3|\xi_0|}} \int_0^{|\partial\Omega|}
    (\kappa(s)+\kappa_0)_+^{1/2}\,ds.
  \end{equation}
\end{theorem}

There is an important difference between Theorems \ref{main1} and
\ref{main2}. If $\mathcal O$ is a bounded domain we can consider both
the interior problem $\Omega=\mathcal O$ and the exterior problem
$\Omega=\R^2\setminus \mathcal O$. Now if $b_0>\Theta_0 b'$ it follows
from \eqref{eq:main} that the leading order terms of $N(h b_0,P_h)$
for both problems coincide  (provided the magnetic fields coincide
on the boundary). This is no longer true if $B$ is constant and
$b_0=\Theta_0 B$. Indeed, the asymptotics are 'complementary' in the
following sense: for the interior problem $N(h\Theta_0 B,P_h)$ is, up
to leading order, determined by the convex part of the boundary (where
the curvature of $\mathcal O$ is positive) and for the exterior
problem $N(h\Theta_0 B,P_h)$ is determined by the concave part (where
the curvature of the obstacle $\mathcal O$ is negative). This
observation is in the same spirit as the considerations on spectral
duality in \cite{HoSm}.

\begin{remark} 
  Let us mention an immediate generalization of Theorems \ref{main1},
  \ref{main2} and their proofs. If $\partial\Omega$ has finitely many
  connected components and if one imposes on each of them either
  Dirichlet or Neumann conditions, then a formula similar to
  \eqref{eq:main} holds, but the integration is restricted to the
  Neumann components. Moreover, it is enough that only the boundary of
  the Neumann components is $C^4$, and also that the assumption
  $\mathbf A\in C^2$ holds only in a neighborhood of those
  components.
\end{remark}

Moreover, it would be desirable to remove the assumption of smoothness of
the boundary. If $\partial\Omega$ is \emph{piecewise} smooth one can
probably use the methods from \cite{Bon}.

%%%%%%%%%%%%%%%%%%%%%%%%%%%%%%%%%%%%%%%%%%%%%%%%%%%%%%%%%%%%%%%%%%%%%%%%%%%%%%%

\subsection{Outline of the paper}

The proofs of Theorems \ref{main1} and \ref{main2} are technical but
the main idea is rather simple. To show \eqref{eq:main} we localize
the problem, following \cite{HeMo}, to a tubular neighborhood of the
boundary of normal size $h^{3/8}$ and cut this into boxes of
tangential size $h^{3/8}$. In each of these boxes we approximate the
magnetic field by a constant one, see Subsection
\ref{sec:bdryest}. This reduces the problem to the analysis of the
model problem of an operator with constant field in a rectangle with
Neumann boundary conditions on one edge and Dirichlet boundary
conditions elsewhere. For this operator we cannot separate variables
but it turns out (Subsection \ref{sec:strip}) that its spectral
counting function is fairly close to the one of the operator on an
infinite half-cylinder. The latter problem  is treated in Subsection
\ref{sec:halfcylinder}. The reduction to the model problem and a
careful estimate of the remainder is achieved in Subsection
~\ref{sec:bracketing}. In Subsection \ref{sec:proofmain} we complete
the proof of Theorem \ref{main1}. We will even obtain a remainder
estimate.

The proof of Theorem \ref{main2} follows a similar pattern and we will
be rather succinct there. The analysis of the model problem is,
however, significantly more difficult, see Subsection
\ref{sec:halfcylinder2}.

On a technical level we mention that classical Dirichlet-Neumann
bracketing is not possible in our situation, since additional Neumann
boundary conditions would produce too many additional
eigenvalues. This difficulty is overcome in \cite{CV} by a
localization technique based on the IMS formula. We emphasize once
more that the paper \cite{CV} concerns Dirichlet boundary conditions,
so that the boundary effects of our Theorems \ref{main1} and
\ref{main2} were not present there.

%%%%%%%%%%%%%%%%%%%%%%%%%%%%%%%%%%%%%%%%%%%%%%%%%%%%%%%%%%%%%%%%%%%%%%%%%%%%%%%

\subsection{Acknowledgements}
The author wishes to thank Prof. B. Helffer for the invitation to
Orsay and numerous fruitful discussions. He is also grateful to
S. Fournais and A. Hansson for useful remarks. Financial support
through the ESF Scientific Programme in Spectral Theory and Partial
Differential Equations (SPECT) as well as through the European
Research Network ``Postdoctoral Training Program in Mathematical
Analysis of Large Quantum Systems'' (Contract Number
HPRN-CT-2002-00277) is gratefully acknowledged.

%%%%%%%%%%%%%%%%%%%%%%%%%%%%%%%%%%%%%%%%%%%%%%%%%%%%%%%%%%%%%%%%%%%%%%
%%%%%%%%%%%%%%%%%%%%%%%%%%%%%%%%%%%%%%%%%%%%%%%%%%%%%%%%%%%%%%%%%%%%%%

\section{Auxiliary material}\label{sec:auxiliary}

\subsection{A family of ordinary differential
  operators}\label{sec:ode}

For any $\xi\in\R$ we consider the operator
\begin{equation}\label{eq:modelop}
  -\frac{d^2}{dt^2}+(\xi+t)^2 \qquad\text{in } L_2(\R_+)
\end{equation}
with a Neumann boundary condition at the origin and denote its lowest
eigenvalue by $\mu(\xi)$. (Here we use the convention of \cite{FoHe1};
we note that in \cite{HeMo} $\xi$ is replaced by $-\xi$.)

The dependence of $\mu(\xi)$ on $\xi$ has been studied in \cite{DaHe}
(see also \cite{Bol}) where the following facts were established: The
function $\mu$ is smooth and satisfies $\lim_{\xi\to-\infty}\mu(\xi)=
1$, $\lim_{\xi\to+\infty}\mu(\xi)= +\infty$. There exists a
$\xi_0\in(-1,0)$ such that $\mu$ is strictly decreasing in
$(-\infty,\xi_0)$ and strictly increasing in
$(\xi_0,\infty)$. Moreover, at $\xi_0$ it has a non-degenerate minimum
and one has
\begin{equation*}
  \Theta_0:=\mu(\xi_0)=\xi_0^2.
\end{equation*}
%We define the constant $\mathcal C_1$ (numerically $\mathcal
%C_1=0.25\dots$) by
%\begin{equation*}
%  \mu''(\xi_0)=6\mathcal C_1 |\xi_0|.
%\end{equation*}

In view of these facts it is possible to introduce two inverse
functions $\nu_+:[\Theta_0,\infty)\to [\xi_0,\infty)$,
    $\nu_-:[\Theta_0,1)\to (-\infty,\xi_0]$ satisfying
\begin{equation*}
  \mu\circ\nu_\pm = \text{id}, \qquad
  \nu_+\circ\mu |_{[\xi_0,\infty)} = \text{id}, \qquad
  \nu_-\circ\mu |_{(-\infty,\xi_0]} = \text{id}.
\end{equation*}
At some points below it will be technically convenient to extend the
functions $\nu_\pm$ by $\xi_0$ to the interval $[0,\Theta_0)$.

Recalling that the minimum of $\mu$ is non-degenerate one easily
establishes

\begin{lemma} 
  For any $0<\epsilon\leq 1$ there exists a constant $C>0$
  such that for all $0\leq\beta \leq\beta'\leq 1-\epsilon$ one has
  \[
  0 \leq \nu_+(\beta')-\nu_+(\beta)\leq C\sqrt{\beta'-\beta}, \qquad
  0 \leq \nu_-(\beta)-\nu_-(\beta')\leq C\sqrt{\beta'-\beta}.
  \]
\end{lemma}

As a consequence one obtains the useful estimate
\begin{equation}\label{eq:hoelder}
	\nu_+(\beta')-\nu_-(\beta')\leq 
	\nu_+(\beta)-\nu_-(\beta)+ 2C\sqrt{\beta'-\beta}.
\end{equation}
Finally, we denote by $\mu_1(\xi)$ the \emph{second} eigenvalue of the
operator \eqref{eq:modelop} and put
\begin{equation*}
  \Theta_1 := \inf_{\xi\in\R} \mu_1(\xi).
\end{equation*}
Numerically, one finds $\Theta_1=2.63\dots$.
\footnote{The author would like to thank V. Bonnaillie-No\"el for this
  calculation.}
Below we shall only use the following bound on $\Theta_1$, the proof
of which is due to B. Helffer.

\begin{lemma}\label{theta1} 
  One has the inequality $\Theta_1>1$.
\end{lemma}

\begin{proof} 
  Denote by $\lambda(\xi)$ the first eigenvalue of the operator
  \eqref{eq:modelop} with a \emph{Dirichlet} boundary condition at the
  origin. Then by Sturm-Liouville theory $\mu_1(\xi)>\lambda(\xi)$ for
  all $\xi\in\R$. Moreover, the variational principle implies that
  \begin{equation}\label{eq:prooftheta1}
    \lambda(\xi)\geq
    \inf\sigma\left(-\frac{d^2}{dt^2}+(\xi+t)^2\right),
  \end{equation}
  where the operator on the RHS is defined in $L_2(\R)$. Using
  translation invariance and the well-known result for the harmonic
  oscillator one finds that the RHS of \eqref{eq:prooftheta1} equals
  1.
\end{proof}

%%%%%%%%%%%%%%%%%%%%%%%%%%%%%%%%%%%%%%%%%%%%%%%%%%%%%%%%%%%%%%%%%%%%%%%%%%%%%%%

\subsection{Boundary coordinates}\label{sec:bdrycoord}

Here we would like to recall the definition of coordinates near the
boundary of $\Omega$. Recall that we assume that $\partial\Omega$ is
connected, $C^4$-smooth and of length
\begin{equation*}
  \mathcal L := |\partial\Omega|.
\end{equation*}
Let $\gamma:\R/\mathcal L\Z\to\partial\Omega$ be a parametrization of 
the boundary with $|\gamma'|\equiv 1$ and let $\nu(s)$ be the interior
unit normal at the point $\gamma(s)$. The parametrization can be
chosen such that $\det(\gamma',\nu)\equiv1$, so that the curvature
$\kappa$ is given by
$\kappa(s)=\langle\gamma''(s),\nu(s)\rangle$.

It is well-known that for sufficiently small $t_0>0$ the map
$\Phi:\R/\mathcal L\Z\times(0,t_0)\to\Omega$,
\begin{equation*}
  \Phi(s,t):=\gamma(s)+t\nu(s),  
\end{equation*}
defines a diffeomorphism between $\R/\mathcal L\Z\times(0,t_0)$ and
its image
\begin{equation*}
  \Phi(\R/\mathcal L\Z\times(0,t_0))= 
  \{x\in\Omega: \dist(x,\Omega)<t_0\} =: \Omega_{t_0}.
\end{equation*}
We now fix a constant $\tilde\kappa\in\R$. (For the proof of Theorem
\ref{main1} it will suffice to take $\tilde\kappa=0$.) Denoting
\begin{equation}\label{eq:defa}
  a(s,t) := 1-t\kappa(s), \qquad (s,t)\in\R/\mathcal L\Z\times(0,t_0),
\end{equation}
and
\begin{equation}\label{eq:defakappa}
  a_{\tilde\kappa}(t):=1-t\tilde\kappa, \qquad t\in\R_+,
\end{equation}
we define for any $u\in L_2(\Omega_{t_0})$ 
\begin{equation}\label{eq:uv} 
  v(s,t) := (a(s,t)/a_{\tilde\kappa}(t))^{1/2} u(\Phi(s,t)),
  \qquad (s,t)\in\R/\mathcal L\Z\times(0,t_0).
\end{equation}
This induces a unitary operator from $L_2(\Omega_{t_0})$ to
$L_2((0,\mathcal L)\times (0,t_0), a_{\tilde\kappa}dsdt)$. As in the
case $\tilde\kappa=0$ considered in \cite{HeMo} (Appendix B) one finds
that if $u\in\mathcal{D}[q_h]$ and $\supp
u\subset\overline{\Omega_{t_0}}$ then
\begin{equation}\label{eq:newcoord}
  \begin{split}
    q_h[u] = 
    & \int_0^\mathcal L \int_0^{t_0}
    (a^{-2}|(hD_s-\tilde{A_1})v|^2+|(hD_t-\tilde{A_2})v|^2+ 
    h^2 W_{\tilde\kappa}|v|^2)\,a_{\tilde\kappa}\, dsdt \\
    & - \frac{h^2}{2}\int_0^\mathcal L (\kappa-\tilde\kappa)
    |v(.,0)|^2\,ds,
  \end{split}
\end{equation}
where
\begin{equation*}
  W_{\tilde\kappa}(s,t) :=
  -\frac{(\kappa(s)-\tilde\kappa)(\kappa(s)+\tilde\kappa(1-2t\kappa(s)))}
  {4a_{\tilde\kappa}(t)^2a(s,t)^2}
  -\frac{t\kappa''(s)}{2a(s,t)^3}
  -\frac{5t^2\kappa'(s)^2}{4a(s,t)^4}.
\end{equation*}
We do not give the expression for $\mathbf{\tilde A}=(\tilde{A_1},
\tilde{A_2})^T$ but note only that
\begin{equation*}
  \tilde B(s,t) :=
  \partial_s \tilde{A_2}(s,t) - \partial_t \tilde{A_1}(s,t) = 
  a(s,t) B(\Phi(s,t)).
\end{equation*}

%%%%%%%%%%%%%%%%%%%%%%%%%%%%%%%%%%%%%%%%%%%%%%%%%%%%%%%%%%%%%%%%%%%%%%%%%%%%%%%
%%%%%%%%%%%%%%%%%%%%%%%%%%%%%%%%%%%%%%%%%%%%%%%%%%%%%%%%%%%%%%%%%%%%%%%%%%%%%%%
%%%%%%%%%%%%%%%%%%%%%%%%%%%%%%%%%%%%%%%%%%%%%%%%%%%%%%%%%%%%%%%%%%%%%%%%%%%%%%%

\section{Proof of Theorem \ref{main1}}

\subsection{The model operator on a
  half-cylinder}\label{sec:halfcylinder} 

We fix $B,S>0$ and consider the operator
$\tilde{P}_h^{S,B}=(h\mathbf{D}-B\mathbf{A}_0)^2$ in
$L_2((0,S)\times\R_+)$ with periodic boundary conditions at
$s\in\{0,S\}$ and Neumann boundary conditions at $t=0$. Here and in
the sequel
\begin{equation}\label{eq:a0}
	\mathbf{A}_0(s,t):=(-t,0)^T.
\end{equation}

\begin{remark} Since we impose periodic boundary conditions, we
  actually work on the half-cylinder $\R/S\Z\times\R_+$. Since this is
  a not a simply connected manifold, the magnetic field alone does not
  determine the operator (up to unitary equivalence), but one also
  needs to specify the circulation of $\mathbf{A}$ around the boundary
  $\R/S\Z\times\{0\}$.
\end{remark}

We shall use the following notation for a self-adjoint and lower
semibounded operator $T$. If $E_T(\Lambda)$, $\Lambda\subset\R$, is
the spectral measure associated with $T$, we put
\[
N(\lambda,T):= \dim\ran E_T\left((-\infty,\lambda)\right), \qquad
\lambda\in\R.
\]
If the spectrum of $T$ below $\lambda$ is discrete, then
$N(\lambda,T)$ coincides with the number of eigenvalues (counting
multiplicities) below $\lambda$.

\begin{lemma}\label{halfcylinder} Let $\lambda<hB$. Then
  \[ 
  N(\lambda,\tilde{P}_h^{S,B}) = 
  \#\left(\Z\cap\frac{B^{1/2}S}{h^{1/2}2\pi}
  \left(\nu_-\left(h^{-1}B^{-1}\lambda\right),
  \nu_+\left(h^{-1}B^{-1}\lambda\right)\right)\right).
  \]
\end{lemma}

Recall that the functions $\nu_\pm$ are extended by $\xi_0$ to
  $[0,\Theta_0)$. In the statement of the lemma we use the notational
  convention that $\left(\nu_-\left(\beta\right),
  \nu_+\left(\beta\right)\right)=(\xi_0,\xi_0)=\emptyset$ if
  $\beta\leq\Theta_0$.

\begin{proof} By separation of variables the operator
  $\tilde{P}_h^{S,B}$ in $L_2((0,S)\times\R_+)$ is unitarily
  equivalent to the direct sum
  \begin{equation*}
    \sum_{n\in\Z}\oplus 
    \left(-h^2\frac{d^2}{dt^2}+\left(2\pi nhS^{-1} +Bt\right)^2
    \right)
    \qquad \text{in }
    \sum_{n\in\Z}\oplus L_2(\R_+)
  \end{equation*}
  (with Neumann boundary conditions at the origin). Applying the
  dilation $\tau=h^{-1/2}B^{1/2}t$ we obtain the unitary equivalence
  \begin{equation*}
    -h^2\frac{d^2}{dt^2}+\left(2\pi nhS^{-1} +Bt\right)^2 \cong 
    hB\left(-\frac{d^2}{d\tau^2}+ 
    \left(2\pi nh^{1/2}B^{-1/2}S^{-1} +\tau\right)^2 \right).    
  \end{equation*}
  By the facts mentioned in Subsection \ref{sec:ode} (particularly
  Lemma \ref{theta1}) we conclude that for $\lambda<hB$
  \begin{equation*}
    \begin{split}
      N(\lambda,\tilde{P}_h^{S,B}) 
      & = \sharp\{n\in\Z:\ \mu(2\pi nh^{1/2}B^{-1/2}S^{-1})
      <h^{-1}B^{-1}\lambda \} \\
      & = \sharp\left(\Z\cap
      (2\pi)^{-1}h^{-1/2}B^{1/2}S \left(\nu_-(h^{-1}B^{-1}\lambda),
      \nu_+(h^{-1}B^{-1}\lambda)\right)\right),
    \end{split} 
  \end{equation*}
  which is what we claimed.
\end{proof}

We note that the proof shows that $N(hB,\tilde{P}_h^{S,B})
=+\infty$. Moreover, we easily deduce from Lemma \ref{halfcylinder}
that for all $b_0<B$
\begin{equation}\label{eq:halfcylinderestimate}
  \left|h^{1/2}N(hb_0,\tilde{P}_h^{S,B}) -
  \frac{B^{1/2}S}{2\pi} \left(\nu_+(B^{-1}
  b_0)-\nu_-(B^{-1}b_0)\right)\right|
  \leq h^{1/2}.
\end{equation}
The relation \eqref{eq:halfcylinderestimate} is in formal accordance
with Theorem \ref{main1} since
\begin{equation*}
  \frac{B^{1/2}S}{2\pi} \left(\nu_+(B^{-1}b_0)-\nu_-(B^{-1}b_0)\right) 
  = \frac1{2\pi} \iint_{\{(s,\xi):\ B\mu(\xi)<b_0\}} B^{1/2}\, dsd\xi.
\end{equation*}

%%%%%%%%%%%%%%%%%%%%%%%%%%%%%%%%%%%%%%%%%%%%%%%%%%%%%%%%%%%%%%%%%%%%%%%%%%%%%%%

\subsection{The model operator on a Dirichlet strip}\label{sec:strip}

We fix $S,T,B>0$ and consider the operator
$P_h^{S,T,B}=(h\mathbf{D}-B\mathbf{A}_0)^2$ in $L_2((0,S)\times
(0,T))$ with Neumann boundary conditions on $t=0$ and Dirichlet
boundary conditions on the remaining part of the boundary. Recall that
$\mathbf A_0$ was defined in \eqref{eq:a0}. Our goal is to compare the
eigenvalue counting function for $P_h^{S,T,B}$ with that of
$\tilde{P}_h^{S,B}$.

\begin{proposition}\label{strip} For all $B, T, S>0$ and all
  $\lambda<hB$ one has
  \begin{equation*}
    N(\lambda,P_h^{S,T,B}) \leq N(\lambda,\tilde{P}_h^{S,B}).
  \end{equation*}
  Moreover, there exists a $C>0$ such that for all $B, T, S>0$, all
  $\delta\in (0,S/2]$ and all $\lambda\leq hB$
  \begin{equation*}
    N(\lambda,P_h^{S,T,B}) \geq
    \frac12
    N(\lambda-Ch^2(\delta^{-2}+T^{-2}),\tilde{P}_h^{2(S-\delta),B}).
  \end{equation*}
\end{proposition}

\begin{proof} 
  The extension by zero of a function in the form domain of
  $P_h^{S,T,B}$ lies in the form domain of $\tilde P_h^{S,B}$, and the
  values of both forms coincide for such a function. Hence the first
  assertion follows immediately by the variational principle.

  To prove the lower bound we follow the ideas of \cite{CV}. For any
  $0<\delta\leq S/2$ we choose a smooth partition of unity on
  $\R/2(S-\delta)\Z$,
\[
(\varphi_1^\delta)^2+(\varphi_2^\delta)^2\equiv1 
\qquad\text{ on } \R/2(S-\delta)\Z,
\]
such that
\[
\supp\varphi_1^\delta \subset [0,S], \qquad 
\supp\varphi_2^\delta \subset [S-\delta,2S-\delta], \qquad 
\sum_{i=1}^2|(\varphi_i^\delta)'|^2\leq c_1\delta^{-2}. 
\]
The constant $c_1>0$ can be chosen independently of
$S,\delta$. Similarly, for each $T>0$ let $\psi_0$, $\psi_1$ fulfill
\begin{equation}\label{eq:psi1}
	(\psi_0^T)^2+(\psi_1^T)^2\equiv 1\qquad\text{ on } \R_+
\end{equation}
and
\begin{equation}\label{eq:psi2}
	\supp\psi_0^T\subset [T/2,+\infty), \qquad \supp\psi_1^T\subset [0,T], 
	\qquad \sum_{i=0}^1|(\psi_i^T)'|^2\leq c_2 T^{-2} 
\end{equation}
with $c_2>0$ independent of $T$. Finally put
$\chi_i^{\delta,T}(s,t):=\varphi_i^\delta(s)\psi_1^T(t)$, $i=1,2$, and
$\chi_0^{\delta,T}(s,t):=\psi_0^T(t)$.

Let $u$ be in the form domain of $\tilde P_h^{2(S-\delta),B}$. Then from
the IMS formula with $I^\delta:=(0,2(S-\delta))$ we get
\begin{equation}\label{eq:stripims}\begin{split}
	& \int_{I^\delta\times\R_+} |(h\mathbf{D}-B\mathbf{A}_0)u|^2\,dx \\
	& \qquad = \sum_{i=0}^2 \int_{I^\delta\times\R_+} |(h\mathbf{D}-B\mathbf{A}_0)\chi_i^{\delta,T} u|^2\,dx
	- h^2 \sum_{i=0}^2 \||\nabla\chi_i^{\delta,T}| u\|^2 \\
	& \qquad \geq \sum_{i=0}^2
	\int_{I^\delta\times\R_+} |(h\mathbf{D}-B\mathbf{A}_0)\chi_i^{\delta,T} u|^2\,dx
	- c_3h^2(\delta^{-2}+T^{-2}) \|u\|^2.
\end{split} \end{equation}
The function $\chi_1^{\delta,T}u$ belongs to the form domain of
$P_h^{S,T,B}$ and, since $u$ is periodic, the function
$\chi_2^{\delta,T}u$ belongs to the form domain of the operator
$\tau_{S-\delta}P_h^{S,T,B}\tau_{S-\delta}^*$ in
$L_2((S-\delta,2S-\delta)\times(0,T))$, where $\tau_{S-\delta}$
denotes translation by $S-\delta$ with respect to the variable $s$. Of
course, $\tau_{S-\delta}P_h^{S,T,B}\tau_{S-\delta}^*$ is unitarily
equivalent to $P_h^{S,T,B}$. Finally, $\chi_0^{\delta,T}u$ belongs to
the form domain of the operator $\tilde P_{h,0}^{2(S-\delta),B} :=
(h\mathbf{D}-B\mathbf{A}_0)^2$ in $L_2(I^\delta\times (T/2,\infty))$
with Dirichlet boundary conditions at $t=T/2$ and periodic boundary
conditions at $s\in\{0,2(S-\delta)\}$. (We do not reflect the
dependence on $T$ in our notation, for operators with different
$T$ are indeed unitarily equivalent.) Hence we deduce from
\eqref{eq:stripims} by the variational principle
\[ 
N(\lambda-c_3h^2(\delta^{-2}+T^{-2}),\tilde{P}_h^{2(S-\delta),B}) \leq 
2 N(\lambda,P_h^{S,T,B}) + N(\lambda,\tilde P_{h,0}^{S,B}).
\]
From an inequality similar to \eqref{eq:dirichletineq} with $\Omega$
replaced by $\R/2(S-\delta)\Z\times\R_+$ we infer that $\tilde
P_{h,0}^{2(S-\delta),B}\geq hB$. Hence $N(\lambda,\tilde
P_{h,0}^{S,B})=0$ if $\lambda\leq hB$ and the proposition is proved.
\end{proof}

%%%%%%%%%%%%%%%%%%%%%%%%%%%%%%%%%%%%%%%%%%%%%%%%%%%%%%%%%%%%%%%%%%%%%%%%%%%%%%%
%%%%%%%%%%%%%%%%%%%%%%%%%%%%%%%%%%%%%%%%%%%%%%%%%%%%%%%%%%%%%%%%%%%%%%%%%%%%%%%

\subsection{Estimates near the boundary}\label{sec:bdryest}

Now we consider a general domain $\Omega\subset\R^2$ such that
$\partial\Omega$ is bounded, connected and $C^4$-smooth. We will
approximate the quadratic form $q_h$ locally near the boundary by a
quadratic form corresponding to a constant magnetic field. For this we
use the boundary coordinates (as well as the notation) introduced in
Subsection \ref{sec:bdrycoord}.
\emph{Throughout this section we will assume that
  $\tilde\kappa=0$.}

Let $T\in (0,t_0]$ and $S\in(0,\mathcal L)$ (below $T, S$ will depend
on $h$ and tend to $0$ as $h\rightarrow 0$). We are interested
in $u\in\mathcal{D}[q_h]$ such that the corresponding $v$,
defined in \eqref{eq:uv}, satisfies
\begin{equation}\label{eq:suppv} 
  \supp v\subset [0,S]\times[0,T]. 
\end{equation}
First we use a gauge transformation to make the field on
$[0,S]\times[0,T]$ 'almost' constant. Indeed, recalling that $\mathbf
A\in C^2(\overline\Omega,\R^2)$, one readily obtains

\begin{lemma}\label{gauge} 
  There exists a constant $C>0$ such that for all $S\in(0,\mathcal
  L)$, $\tilde S\in[0,S]$ there exists a function $\phi\in
  C^2([0,S]\times[0,t_0])$ such that
  \begin{equation*}
    \mathbf{\tilde A}(s,t)-\mathbf\nabla\phi(s,t)
    =(-\tilde B t+\beta(s,t),0)^T, \qquad (s,t)\in[0,S]\times[0,t_0],
  \end{equation*}
  where $\tilde B:=\tilde B(\tilde S,0)$ and for any $0<T\leq t_0$
  \begin{equation}\label{eq:gaugeremainder}
    \sup_{(s,t)\in[0,S]\times[0,T]}|\beta(s,t)| \leq C(S^2+T^2).
  \end{equation}
\end{lemma}

If $u$ is supported in a small subset near the boundary the previous
lemma allows us to express $q_h[u]$, up to a small error, via a
quadratic form corresponding to a constant magnetic field.

\begin{lemma}\label{bdrycoord}
  There exists a constant $C>0$ such that for all $S\in(0,\mathcal
  L)$, $\tilde S\in[0,S]$, $T\in(0,t_0]$, $\epsilon\in(0,1]$
  satisfying $\epsilon\geq CT$ and for all $u\in\mathcal{D}[q_h]$ such
  that the corresponding $v$ satisfies \eqref{eq:suppv} one has
  \begin{equation*}
    \begin{split}
      & \left| q_h[u]- 
      \|(h\mathbf{D}-\tilde{B}\mathbf{A}_0)e^{-i\phi/h}v\|^2 \right| \\
      & \qquad \leq 
      \epsilon \|(h\mathbf{D}-\tilde{B}\mathbf{A}_0)e^{-i\phi/h}v\|^2 + 
      C\epsilon^{-1}\left((S^2+T^2)^2+h^2\right)\|e^{-i\phi/h}v\|^2.
    \end{split}  
  \end{equation*}
  Here $\tilde B:=\tilde B(\tilde S,0)$ and $\phi$ is the function
  from Lemma \ref{gauge}.
\end{lemma}

\begin{proof} 
  We write $w:=e^{-i\phi/h}v$. In view of \eqref{eq:newcoord} we decompose
  \begin{equation*}
    q_h[u]- \|(h\mathbf{D}-\tilde{B}\mathbf{A}_0)w\|^2 = I_1+I_2+I_3,
  \end{equation*}
  where we define
  \begin{equation*}
    \begin{split}
      I_1 & :=  \int_0^\mathcal L \int_0^{t_0}
      \left(a^{-2}|(hD_s-\tilde{A_1})v|^2+|(hD_t-\tilde{A_2})v|^2\right)
      \,dsdt \\
      & \qquad - \|(h\mathbf{D}-\tilde{B}\mathbf{A}_0)w\|^2,\\
      I_2 & := h^2 \int_0^\mathcal L \int_0^{t_0} W_0|v|^2\,dsdt, \\
      I_3 & := -\frac{h^2}{2}\int_0^\mathcal L \kappa |v(.,0)|^2\,ds.
    \end{split}
  \end{equation*}
  We begin with the (easier) terms $I_2$ and $I_3$. Since $W_0$ is
  bounded, we obtain that for some constant $c_1>0$
  \begin{equation}\label{eq:i2}
    |I_2|\leq c_1 h^2 \|v\|^2.
  \end{equation}
  Moreover, for any $s\in[0,L]$ one has
  \begin{equation*}
    \begin{split}
      |w(s,0)|^2  
      & = - 2\text{Re }\int_0^T \frac{\partial}{\partial t}w(s,t) 
      \overline{w(s,t)}\,dt \\
      & \leq \int_0^T \left(\epsilon_1|D_tw(s,t)|^2 + 
      \epsilon_1^{-1}|w(s,t)|^2\right)\,dt.
    \end{split} 
  \end{equation*}
  Since $\kappa$ is bounded one easily concludes that there is a
  constant $c_2>0$ such that
  \begin{equation}\label{eq:i3}
    |I_3| \leq \epsilon \|hD_t w\|^2 + c_2\epsilon^{-1}h^2 \|w\|^2
  \end{equation}
  for any $\epsilon>0$. Now we turn to the term $I_1$. First we note
  that by Lemma ~\ref{gauge} one has
  \begin{equation*}
    I_1= \|a^{-1}(hD_s+\tilde{B}t-\beta)w\|^2 - \|(hD_s+\tilde{B}t)w\|^2.
  \end{equation*}
  We use that for some constant $c_3 >0$
  \begin{equation*}
    |a^{-2} - 1|\leq c_3 T \qquad\text{on } [0,S]\times[0,T],  
  \end{equation*}
  and hence for all $\epsilon_2>0$
  \begin{equation*}
    \begin{split}
      |I_1| 
      & \leq (1+c_3T)\|(hD_s+\tilde{B}t-\beta)w\|^2 
      - \|(hD_s+\tilde{B}t)w\|^2 \\
      & \leq ((1+c_3T)(1+\epsilon_2)-1) \|(hD_s+\tilde{B}t)w\|^2 
      + (1+c_3T)(1+\epsilon_2^{-1})\|\beta w\|^2.
    \end{split} 
  \end{equation*}
  In particular, if $\epsilon_2\geq T$ then
  $(1+c_3T)(1+\epsilon_2)-1\leq c_4\epsilon_2$. Recalling
  \eqref{eq:gaugeremainder} we obtain
  \begin{equation}\label{eq:i1}
    |I_1| \leq c_4\epsilon_2 \|(hD_s+\tilde{B}t)w\|^2 + 
    c_5 (S^2+T^2)^2 (1+\epsilon_2^{-1}) \|w\|^2.
  \end{equation}
  The assertion now follows easily by setting $c_4\epsilon_2=\epsilon$
  and summing \eqref{eq:i2}, \eqref{eq:i3}, \eqref{eq:i1}.
\end{proof}

%%%%%%%%%%%%%%%%%%%%%%%%%%%%%%%%%%%%%%%%%%%%%%%%%%%%%%%%%%%%%%%%%%%%%%%%%%%%%%%

\subsection{Bracketing}\label{sec:bracketing}

Now we combine Proposition \ref{strip} and Lemma \ref{bdrycoord} in
order to obtain a two-sided estimate of $N(hb_0,P_h)$ in terms of the
spectral counting functions of operators on a half-cylinder with
constant magnetic fields.

For $N\in\N$ put 
\begin{equation}\label{eq:s} 
  S:=\frac{\mathcal L}N \qquad 
  \text{and} \qquad
  s_n:=nS, \, n=0,\ldots,N. 
\end{equation}
We choose
\begin{equation*}
  N=[h^{-3/8}],  
\end{equation*}
and note that $S=\mathcal O(h^{3/8})$. Then the core of Theorem
\ref{main1} is contained in

\begin{proposition}\label{bracketing} 
  Let $b_0<b$. Under the above hypotheses there exists a constant
  $C>0$ such that for any $0<h\leq C^{-1}$, $\delta\in(0,S/2]$ and
  $\tilde S_n\in [s_{n-1},s_n]$, $n=1,\ldots,N$, one has
  \begin{equation*}\label{eq:bracketing}
    \begin{split}
      \frac12 \sum_{n=1}^N N(hb_0 - C h^2\delta^{-2}, 
      \tilde P_h^{2(S-\delta),\tilde{B}_n}) 
      & \leq N(hb_0,P_h) \\
      & \leq
      \sum_{n=1}^N N(hb_0 + C h^2\delta^{-2}, 
      \tilde P_h^{S+2\delta,\tilde{B}_n})
    \end{split}
  \end{equation*}
  where $\tilde B_n:=\tilde B(\tilde S_n,0)$.
\end{proposition}

\begin{proof}
  We begin with the proof of the lower bound. We apply Lemma
  \ref{bdrycoord} with
  \begin{equation*}\label{eq:choice} 
    T=h^{3/8}, \qquad \epsilon=h^{1/4}
  \end{equation*}
  not only on $[0,S]$, but on any $[s_{n-1},s_n]$. (It is evident from
  the proof of that lemma that the constants there can be chosen
  independently of $n$.) Moreover, recall that
  $S=\mathcal{O}(h^{3/8})$. It follows that, for some constant $c_1>0$
  and for all $u\in\mathcal{D}[q_h]$ which vanish on
  \begin{equation*}
    \{x\in\Omega :\ t(x)\geq T\}\cup 
    \bigcup_{n=1}^N\{x\in\Omega :\  0\leq t(x)\leq T, s(x)=s_n \},
  \end{equation*}
  one has the estimate
  \begin{equation*}
    \begin{split}
      q_h[u] \leq &
      (1+h^{1/4})\sum_{n=1}^N 
      \int_{s_{n-1}}^{s_n} \int_0^T 
      |(h\mathbf{D}-\tilde{B}_n\mathbf{A}_0)e^{-i\phi_n/h}v|^2\,dsdt
      \\
      & + c_1 h^{5/4}\|e^{-i\phi_n/h} v\|^2, 
    \end{split}  
  \end{equation*}
  where $v$ is defined by \eqref{eq:uv}. Hence by the variational principle
  \begin{equation*}
    N(\lambda,P_h) \geq 
    \sum_{n=1}^N N\left(\frac{\lambda-c_1 h^{5/4}}{1+h^{1/4}},
    P_h^{S,T,\tilde{B}_n}\right).    
  \end{equation*}
  We apply Proposition \ref{strip} with $\delta\in (0,S/2]$. Since 
  \begin{equation*}
    \frac{hb_0-c_1 h^{5/4}}{1+h^{1/4}} - Ch^2(\delta^{-2} + T^{-2})
    \geq hb_0 - c_2 h^2 \delta^{-2},
  \end{equation*}
  we obtain
  \begin{equation*}
    N(hb_0,P_h) \geq
    \frac12 \sum_{n=1}^N 
    N(hb_0 - c_2 h^2\delta^{-2}, \tilde P_h^{2(S-\delta),\tilde{B}_n} ).    
  \end{equation*}
  This is the desired lower bound.

  We turn to the proof of the upper bound. We choose for every
  $\delta\in (0,S/2]$ a partition of unity on $\R/\mathcal L\Z$,
  \begin{equation*}
    \sum_{i=1}^N (\varphi_n^\delta)^2 \equiv 1 \text{ on }
    \R/\mathcal L\Z \qquad\text{with } \ 
    \supp\varphi_n^\delta \subset [s_{n-1}-\delta, s_n+\delta] 
  \end{equation*}
  and such that
  \begin{equation*}
    \sum_{i=1}^N |(\varphi_n^\delta)'|^2 \leq c_3\delta^{-2}.
  \end{equation*}
  The constant $c_3>0$ can be chosen independent of $\delta, S, N$.
  Moreover, let $\psi_0^T$, $\psi_1^T$ be as in \eqref{eq:psi1},
  \eqref{eq:psi2}, and for $T\in (0,t_0)$ put
  \begin{equation*}
    \chi_n^{\delta,T}(x) := 
    \varphi_n^\delta(s(x))\psi_1^T(t(x)), \ n=1,\ldots,N, \qquad
    \chi_0^{\delta,T}(x) := \psi_0^T(t(x)).
  \end{equation*}
  By means of the IMS formula we find for all $u\in\mathcal D[q_h]$
  \begin{equation*}
    \begin{split} 
      q_h[u] & = 
      \sum_{n=0}^N q_h[\chi_n^{\delta,T}u] 
      - h^2 \sum_{n=0}^N \||\nabla\chi_n^{\delta,T}| u\|^2 \\
      & \geq \sum_{n=0}^N q_h[\chi_n^{\delta,T}u] 
      - c_4 h^2(\delta^{-2}+T^{-2}) \|u\|^2.
    \end{split}  
  \end{equation*}
  As in the proof of the upper bound choose $T$, $\epsilon$ as in
  \eqref{eq:choice}. Then we obtain from Lemma \ref{bdrycoord} that
  for $n=1,\ldots,N$
  \begin{equation*}
    \begin{split}
      q_h[\chi_n^{\delta,T}u] & \geq \left(1-h^{1/4}\right) 
      \int_{s_{n-1}}^{s_n} \int_0^T 
      |(h\mathbf{D}-\tilde{B}_n\mathbf{A}_0)e^{-i\phi_n/h}\chi_n^{\delta,T}v|^2
      \,dsdt \\
      & \qquad -c_5 h^{5/4} \|e^{-i\phi_n/h}\chi_n^{\delta,T}v\|^2,
    \end{split} 
  \end{equation*}
  where $v$ is related to $u$ by \eqref{eq:uv} and where we write
  $\chi_n^{\delta,T}$ on the RHS instead of
  $\chi_n^{\delta,T}\circ\Phi$. It follows that
  \begin{equation*}
    \begin{split}
      q_h[u] & \geq 
      \left(1-h^{1/4}\right) \sum_{n=1}^N \int_{s_{n-1}}^{s_n} \int_0^T
      |(h\mathbf{D}-\tilde{B}_n\mathbf{A}_0)e^{-i\phi_n/h}\chi_n^{\delta,T}v|^2
      \,dsdt \\
      & \qquad + q_h[\chi_0^{\delta,T}u] - c_6 h^2\delta^{-2} \|u\|^2,
    \end{split} 
  \end{equation*}
  and hence by the variational principle
  \begin{equation*}
    N(\lambda,P_h) \leq 
    \sum_{n=1}^N 
    N\left(\frac{\lambda+c_6 h^2\delta^{-2}}{1-h^{1/4}},
    P_h^{S+2\delta,T,\tilde{B}_n} \right)
    + N(\lambda+c_6 h^2\delta^{-2}, P_{h,0}^T),    
  \end{equation*}
  where $P_{h,0}^T:=(h\mathbf{D}-\mathbf{A})^2$ in
  $L_2(\Omega\setminus\overline{\Omega_{T/2}})$ with Dirichlet
  boundary conditions. Because of the inequality
  \eqref{eq:dirichletineq} we have $N(\lambda,P_{h,0}^T)=0$ for
  $\lambda\leq hb$. Moreover, since
  \begin{equation*}
    \frac{hb_0+c_6 h^2\delta^{-2}}{1- h^{1/4}} 
    \leq hb_0 + c_7 h^2 \delta^{-2},
  \end{equation*}
  the desired upper bound follows from Proposition \ref{strip}.
\end{proof}

%%%%%%%%%%%%%%%%%%%%%%%%%%%%%%%%%%%%%%%%%%%%%%%%%%%%%%%%%%%%%%%%%%%%%%%%%%%%%%%

\subsection{Proof of Theorem \ref{main1}}\label{sec:proofmain}

To complete the proof of our main result we now combine Proposition
\ref{bracketing} with the explicit calculation in Lemma
\ref{halfcylinder} for the half-cylinder. Indeed, instead of
\eqref{eq:main} we prove the stronger (albeit probably not sharp)
estimate
\begin{equation}\label{eq:mainremainder}
  h^{1/2} N(hb_0,P_h) = 
  \frac{1}{2\pi} \int_{\partial\Omega}
  \int_{\{\xi:\ \mu(\xi)<\frac{b_0}{B(x)}\}} 
  B(x)^{1/2} \,d\xi\,ds(x) 
  +\mathcal O(h^{1/16})
\end{equation}
for $b_0<b$. The proofs of the upper and the lower bound in
\eqref{eq:mainremainder} are similar and we only give the latter one.

Let $N$, $S$ be as in Subsection \ref{sec:bracketing} and let $\tilde
S_n\in [s_{n-1},s_n]$ be arbitrary with $\tilde B_n:=\tilde B(\tilde
S_n,0)$. Note that for all sufficiently small $h$ one has $b_0 -C
h\delta^{-2} < b \leq b'\leq \tilde B_n$. Hence, from
\eqref{eq:halfcylinderestimate} and Proposition \ref{bracketing}, for
all $\delta\in(0,S/2]$
\[ \begin{split}
h^{1/2} N(hb_0,P_h) \geq & 
\frac1{2\pi}\sum_{n=1}^N (S-\delta) \tilde{B}_n^{1/2}\times \\
& \quad \times \left(\nu_+\left(\tilde B_n^{-1}(b_0 -C h\delta^{-2})\right)-
	\nu_-\left(\tilde B_n^{-1}(b_0 - C h\delta^{-2})\right)\right) \\
& -\frac{N}{2}h^{1/2}.
\end{split} \]
Now we use \eqref{eq:hoelder}, choose $\delta=h^{7/16}$ and recall that $N=\mathcal O(h^{-3/8})$ to get
\[ 
	h^{1/2} N(hb_0,P_h) \geq 
	\frac{1}{2\pi} \sum_{n=1}^N S \tilde{B}_n^{1/2}
	\left(\nu_+(\tilde B_n^{-1} b_0)-\nu_-(\tilde B_n^{-1} b_0)\right) - c_1 h^{1/16}. 
\]
The main term on the RHS is a Riemannian sum. Recalling that the $\tilde S_n$ were arbitrary, we finally arrive at
\begin{equation*}
  \begin{split}
    & h^{1/2} N(hb_0,P_h) \\
    & \qquad \geq \frac{1}{2\pi}\int_{\partial\Omega}
    B(x)^{1/2}\left(\nu_+\left(\frac{b_0}{B(x)}\right)
    -\nu_-\left(\frac{b_0}{B(x)}\right)\right)\,ds(x) 
    - c_1 h^{1/16} \\
    & \qquad = \frac{1}{2\pi}\int_{\partial\Omega} 
    \int_{\{\xi:\ \mu(\xi)<\frac{b_0}{B(x)}\}} B(x)^{1/2}\,d\xi\,ds(x) 
    - c_1 h^{1/16}.
  \end{split}
\end{equation*}
This is the lower bound of \eqref{eq:mainremainder}. In a similar
fashion one can establish the upper bound, which concludes the proof
of Theorem \ref{main1}.

%%%%%%%%%%%%%%%%%%%%%%%%%%%%%%%%%%%%%%%%%%%%%%%%%%%%%%%%%%%%%%%%%%%%%%%%%%%%%%%
%%%%%%%%%%%%%%%%%%%%%%%%%%%%%%%%%%%%%%%%%%%%%%%%%%%%%%%%%%%%%%%%%%%%%%%%%%%%%%%
%%%%%%%%%%%%%%%%%%%%%%%%%%%%%%%%%%%%%%%%%%%%%%%%%%%%%%%%%%%%%%%%%%%%%%%%%%%%%%%

\section{Proof of Theorem \ref{main2}}

\subsection{The model operator on a
  half-cylinder}\label{sec:halfcylinder2}

For parameters $S,T>0$, $\kappa\in\R$ satisfying
\begin{equation*}
  2|\kappa|T\leq 1,
\end{equation*}
we denote by $\tilde M_h^{S,T,\kappa}$ the self-adjoint operator in
$L_2((0,S)\times (0,T),a_\kappa dsdt)$ associated with the quadratic
form
\begin{equation*} 
  \begin{split}
    \tilde m_h^{S,T,\kappa}[u] & := \int_0^S \int_0^T 
    \left(a_\kappa^{-2} |(hD_s+t-\kappa t^2/2)u|^2 + |hD_tu|^2 \right)
    a_\kappa\,dsdt, \\
    \mathcal D[\tilde m_h^{S,T,\kappa}] & :=
    \{ u \in H^1((0,S)\times(0,T)) :\ 
    u(.,T)=0,\  u(0,.)=u(S,.) \}.
  \end{split}
\end{equation*}
Recall that the function $a_\kappa$ was defined in
\eqref{eq:defakappa}. (We emphasize once more that here and in the
next subsection, $\kappa$ will be a constant and not the curvature.)

Moreover, recall the constant $\mathcal C_1$ from \eqref{eq:c1}. The
goal of this subsection is to prove

\begin{proposition}\label{halfcylinder2}
  Let $D>0$. Then there exist $C,\epsilon>0$ such that for all
  $|\kappa|\leq D$, $|\kappa_0|\leq D$, $S>0$, $0<h\leq\epsilon$ and
  $\epsilon^{-1} \sqrt h |\log h| \leq T\leq\epsilon h^{1/4}$ one has
  \begin{equation*}
    \left| N(h\Theta_0+h^{3/2}\mathcal C_1 \kappa_0, \tilde M_h^{S,T,\kappa}) 
    - h^{-1/4} \frac S{\pi\sqrt{3|\xi_0|}} (\kappa+\kappa_0)_+^{1/2} \right|
    \leq C.
  \end{equation*}
\end{proposition}

For the analysis of the operators $\tilde M_h^{S,T,\kappa}$ we begin
as in the proof of Lemma ~\ref{halfcylinder}. By separation of variables
and a dilation $\tau=h^{-1/2}t$ we obtain the unitary
equivalence
\begin{equation}\label{eq:halfcylinder2decomp}
  \tilde M_h^{S,T,\kappa} \cong
  h \sum_{n\in\Z}\oplus M(2\pi nh^{1/2}S^{-1},
  h^{1/2}\kappa,h^{-1/2}T). 
\end{equation}
Here we define, for parameters
\begin{equation*}
  \xi\in\R, \qquad \alpha\in [-1,1], \qquad L\geq 1, \qquad
  2|\alpha|L\leq 1,
\end{equation*}
the self-adjoint operator $M(\xi,\alpha,L)$ in the Hilbert space
$L_2((0,L),a_\kappa d\tau)$ by the quadratic form
\begin{equation*} 
  \begin{split}
    m(\xi,\alpha,L)[f] & := \int_0^L \left(|f'|^2 + a_\alpha^{-2}
    (\xi+\tau-\alpha \tau^2/2)^2|f|^2\right) a_\alpha \,d\tau, \\
    \mathcal D[m(\xi,\alpha,L)] & :=
    \{ f \in H^1(0,L) :\ f(L)=0 \}.
  \end{split}
\end{equation*}
The proof of Proposition \ref{halfcylinder2} relies on the following
two results, which we take from \cite{HeMo} (Section 11), see also
\cite{FoHe1} (Lemma 5.4). We shall denote the eigenvalues of a
self-adjoint and lower semibounded operator $T$ with compact resolvent
by $\mu_1(T)\leq \mu_2(T)\leq\ldots$, taking multiplicities into account.

\begin{lemma}\label{modeloned1}
  Let $D>0$. Then there exists a $C>0$ such that if $|\alpha|L^2\leq
  D$ then for all $j\in\N$
  \begin{equation*}
    |\mu_j(M(\xi,\alpha,L)) - \mu_j(M(\xi,0,L))| 
    \leq C|\alpha|L^2(1+ \mu_j(M(\xi,0,L))).
  \end{equation*}
\end{lemma}

The previous lemma follows easily by comparing the corresponding
quadratic forms. Using an explicit trial function one can show

\begin{lemma}\label{modeloned2} 
  Let $D>0$. Then there exists a $C>0$ such that for all
  $|\xi-\xi_0|\leq D$, $L\geq C$ there exists a
  $\lambda\in\sigma(M(\xi,\alpha,L))$ with
  \begin{equation}\label{eq:modeloned2}
    \begin{split}
      & \left| \lambda - \Theta_0-3\mathcal C_1|\xi_0|(\xi-\xi_0)^2 + 
      \mathcal C_1\alpha \right| \\
      & \qquad \leq
      C\left(|\xi-\xi_0|^3+|\alpha||\xi-\xi_0|+\alpha^2+e^{-L/C}\right).
    \end{split}
  \end{equation}
\end{lemma}

We will use the following consequence of the two preceding lemmas.

\begin{corollary}\label{modeloned3}
  Let $\beta>0$. Then there exist $\epsilon,\delta, C>0$ such that for
  all $\xi\in\R$, $|\alpha|L^2\leq\epsilon$, $L\geq\epsilon^{-1}$ one
  has:
  \begin{enumerate}
    \item
      If $|\xi-\xi_0|\leq \beta C^{-1}$ then
      \begin{equation*}
	\begin{split}
	  & \left| \mu_1(M(\xi,\alpha,L)) 
	  - \Theta_0-3\mathcal C_1|\xi_0|(\xi-\xi_0)^2 +
	  \mathcal C_1\alpha \right| \\
	  & \qquad \leq
	  C\left(|\xi-\xi_0|^3+|\alpha||\xi-\xi_0|+\alpha^2+e^{-L/C}\right)
	\end{split}
      \end{equation*}
      and
      \begin{equation*}
	\mu_2(M(\xi,\alpha,L)) \geq \Theta_0+\delta.
      \end{equation*}
    \item
      If $|\xi-\xi_0|\geq \beta C^{-1}$ then
      \begin{equation*}
	\mu_1(M(\xi,\alpha,L)) \geq \Theta_0+\delta.
      \end{equation*}
  \end{enumerate}
\end{corollary}

The parameter $\beta$ is introduced for technical reasons which will
become clear in the proof of Proposition \ref{halfcylinder2}.

\begin{proof}[Proof of Corollary \ref{modeloned3}]
  By Lemma \ref{modeloned2} there exist $\epsilon, C>0$ such that if
  $|\xi-\xi_0|\leq \beta C^{-1}$, $|\alpha|L^2\leq\epsilon$ and
  $L\geq\epsilon^{-1}$ then $M(\xi,\alpha,L)$ has an eigenvalue
  $\lambda$ below $\frac12(\Theta_0+\Theta_1)$ satisfying
  \eqref{eq:modeloned2} with the constant $C$. (We could of course
  choose $\epsilon=C^{-1}$ or $\epsilon=\beta C^{-1}$, but later it
  will be useful to keep them separated.) Note that by the variational
  principle we have
  \begin{equation}\label{eq:modelonedvar}
    \mu_j(M(\xi,0,L)) \geq \mu_j(M(\xi)),
  \end{equation}
  where $M(\xi)$ denotes the operator studied in Subsection
  \ref{sec:ode} (corresponding to $\alpha=0$ and $L=\infty$). Hence by
  Lemma \ref{modeloned1} we find that, after decreasing $\epsilon$ if
  necessary, one has for all $|\xi-\xi_0|\leq \beta C^{-1}$
  \begin{equation*}
    \begin{split}
      \mu_2(M(\xi,\alpha,L))
      & \geq (1-c_1|\alpha|L^2)\mu_2(M(\xi,0,L)) - c_1|\alpha|L^2\\
      & \geq (1-c_1|\alpha|L^2)\mu_2(M(\xi)) - c_1|\alpha|L^2\\
      & \geq (1-c_1|\alpha|L^2)\Theta_1 - c_1|\alpha|L^2\\
      & \geq (\Theta_0+\Theta_1)/2. 
    \end{split}
  \end{equation*}
  In particular, it follows that
  $\lambda=\mu_1(M(\xi,\alpha,L))$. This finishes the proof of the
  first part of the corollary (with
  $\delta\leq\frac12(\Theta_1-\Theta_0)$ arbitrary).

  By the properties of the function $\mu$ recalled in Subsection
  \ref{sec:ode} it is clear that there exists a constant $\delta>0$
  such that for all $|\xi-\xi_0|\geq\beta C^{-1}$ one has
  \begin{equation*}
    \mu(\xi)\geq \Theta_0 +2\delta.
  \end{equation*}
  (We can assume that $\delta\leq\frac12(\Theta_0+\Theta_1)$.)
  Applying again \eqref{eq:modelonedvar} and Lemma ~\ref{modeloned1}
  and decreasing $\epsilon$ if necessary we find for all
  $|\xi-\xi_0|\geq\beta C^{-1}$
  \begin{equation*}
    \begin{split}
      \mu_1(M(\xi,\alpha,L))
      & \geq (1-c_1|\alpha|L^2)\mu_1(M(\xi,0,L)) - c_1|\alpha|L^2\\
      & \geq (1-c_1|\alpha|L^2)\mu(\xi) - c_1|\alpha|L^2\\
      & \geq (1-c_1|\alpha|L^2)(\Theta_0+2\delta) - c_1|\alpha|L^2\\
      & \geq \Theta_0+\delta.
    \end{split}
  \end{equation*}
  This finishes the proof of the second part of the corollary.
\end{proof}

Everything is now in place for the

\begin{proof}[Proof of Proposition \ref{halfcylinder2}.]
  We keep the notation $\epsilon,\delta,C$ for the constants from
  Corollary \ref{modeloned3} corresponding to
  \begin{equation}\label{eq:defbeta}
    \beta := 3\mathcal C_1 |\xi_0|/2,
  \end{equation}
  and we will assume that
  \begin{equation*}
    C\sqrt h |\log h| \leq T \leq \sqrt{\epsilon/D}\, h^{1/4}.
  \end{equation*}
  Then there exists a $h_0>0$ such that for all $0<h\leq h_0$ and all
  $|\kappa|,|\kappa_0|\leq D$ one has
  $h^{-1/2}|\kappa|T^2\leq\epsilon$, $h^{-1/2}T\geq\epsilon^{-1}$ and
  $h^{1/2} \mathcal C_1 \kappa_0 \leq\delta$. Therefore Corollary
  \ref{modeloned3} and the decomposition
  \eqref{eq:halfcylinder2decomp} imply that
  \begin{equation*}
    \begin{split}
      & N(h\Theta_0+h^{3/2}\mathcal C_1 \kappa_0, \tilde
      M_h^{S,T,\kappa})\\ 
      & \qquad = \sharp\{n\in\Z: \
      \mu_1(M(2\pi nh^{1/2}S^{-1},h^{1/2}\kappa,h^{-1/2}T))
      <\Theta_0+h^{1/2}\mathcal C_1 \kappa_0 \}.
    \end{split}
  \end{equation*}
  Noting that $h\kappa^2 + e^{-T/C\sqrt h} \leq h(D^2+1)$ we find from
  Corollary \ref{modeloned3} the estimates
  \begin{equation}\label{eq:modelonedestimate}
    \begin{split}
      & \sharp\{n\in\Z:\ p_+(|2\pi n S^{-1}-h^{-1/2}\xi_0|)<0 \} \\
      & \qquad \leq N(h\Theta_0+h^{3/2}\mathcal C_1 \kappa_0, \tilde
      M_h^{S,T,\kappa}) \\
      & \qquad \leq  \sharp\{n\in\Z:\ |2\pi nS^{-1}-h^{-1/2}\xi_0|\leq
      \beta C^{-1} h^{-1/2},\\
      & \qquad \qquad \qquad p_-(|2\pi n S^{-1}-h^{-1/2}\xi_0|)<0 \}
    \end{split}
  \end{equation}
  where
  \begin{equation*}
    p_{\pm,h}(y) := 3\mathcal C_1|\xi_0| y^2
    - \mathcal C_1h^{-1/2}(\kappa+\kappa_0)
    \pm C(h^{1/2} y^3 + Dy + D^2+1).
  \end{equation*}
  The assertion will follow from the properties of these polynomials
  which we will discuss now briefly.

  We start with $p_{+,h}$. If $\kappa+\kappa_0\leq \mathcal C_1^{-1}
  C(D^2+1) h^{1/2}$ one has $p_{+,h}(y)> 0$ for all $y>0$ and we
  define $y_{+,h}:=0$. On the other hand, if $\kappa+\kappa_0>
  \mathcal C_1^{-1} C(D^2+1) h^{1/2}$, one checks that there is a
  unique zero $y_{+,h}\in(0,\infty)$, and that this satisfies
  \begin{equation}\label{eq:modelonedzero}
    y_{+,h} = 
    (3|\xi_0|)^{-1/2}(\kappa+\kappa_0)_+^{1/2}h^{-1/4} + \mathcal O(1)
  \end{equation}
  as $h\to 0$, where $\mathcal O(1)$ is uniform in $\kappa$,
  $\kappa_0$ (varying in a bounded set). Hence in any case, we obtain
  from \eqref{eq:modelonedestimate} the lower bound
  \begin{equation*}
    \begin{split}
      N(h\Theta_0+h^{3/2}\mathcal C_1 \kappa_0, \tilde
      M_h^{S,T,\kappa})
      & \geq  \sharp\{n\in\Z: |2\pi nS^{-1}-h^{-1/2}\xi_0|< y_{+,h} \}
      \\
      & =  h^{-1/4} \frac S{\pi\sqrt{3|\xi_0|}}
      (\kappa+\kappa_0)_+^{1/2} + \mathcal O(1).
    \end{split}
  \end{equation*}

  Now we turn to the polynomial $p_{-,h}$. First we note that there is
  a zero  $\tilde y_{-,h}\sim 3\mathcal C_1|\xi_0| C^{-1}h^{-1/2}$,
  which however does not lie in the interval
  $[0,\beta C^{-1} h^{-1/2})$ if $h$ is sufficiently small. (This is
  the reason for the choice of $\beta$ in \eqref{eq:defbeta}.) If
  $\kappa+\kappa_0 \leq -\mathcal C_1^{-1} C(D^2+1) h^{1/2}$, $\tilde
  y_{-,h}$ is the only zero in $(0,\infty)$ and we set
  $y_{-,h}:=0$. On the other hand, if $\kappa+\kappa_0 > -\mathcal
  C_1^{-1} C(D^2+1) h^{1/2}$, one checks that there is a unique zero
  $y_{-,h}\in[0,\tilde y_{-,h})$, and that this has the same expansion
  as in \eqref{eq:modelonedzero}. In both cases, we obtain the upper
  bound
  \begin{equation*}
    \begin{split}
      N(h\Theta_0+h^{3/2}\mathcal C_1 \kappa_0, \tilde
      M_h^{S,T,\kappa})
      & \leq  \sharp\{n\in\Z: |2\pi nS^{-1}-h^{-1/2}\xi_0|< y_{-,h}\}
      \\
      & =  h^{-1/4} \frac S{\pi\sqrt{3|\xi_0|}} 
      (\kappa+\kappa_0)_+^{1/2} + \mathcal O(1).
    \end{split}
  \end{equation*}
  This proves the assertion.
\end{proof}

%%%%%%%%%%%%%%%%%%%%%%%%%%%%%%%%%%%%%%%%%%%%%%%%%%%%%%%%%%%%%%%%%%%%%%%%%%%%%%

\subsection{The model operator on a Dirichlet strip}\label{sec:strip2}

We fix $S,T,\kappa$ as in the previous subsection and consider the
operator $M_h^{S,T,\kappa}$ obtained from $\tilde M_h^{S,T,\kappa}$ by
imposing additional Dirichlet boundary conditions at
$s\in\{0,S\}$. More precisely, $M_h^{S,T,\kappa}$ is the self-adjoint
operator in $L_2((0,S)\times (0,T),a_\kappa dsdt)$ associated with the
quadratic form
\begin{equation*} 
  \begin{split}
    m_h^{S,T,\kappa}[u] & := \int_0^S \int_0^T 
    \left(a_\kappa^{-2} |(hD_s+t-\kappa t^2/2)u|^2 + |hD_tu|^2 \right)
    a_\kappa\,dsdt, \\
    \mathcal D[m_h^{S,T,\kappa}] & :=
    \{ u \in H^1((0,S)\times(0,T)) :\ 
    u(.,T)=u(0,.)=u(S,.)=0 \}.
  \end{split}
\end{equation*}
With an argument similar to that in Subsection \ref{sec:strip} one
proves

\begin{proposition}\label{strip2}
  There exists a $C>0$ such that for all $S, T>0$, $\kappa\in\R$ with
  $2|\kappa|T\leq 1$ and all $\lambda\in\R$ and $\delta\in (0,S/2]$
  one has
  \begin{equation*}
    \frac12 N(\lambda-Ch^2\delta^{-2},
    \tilde M_h^{2(S-\delta),T,\kappa})
    \leq N(\lambda,M_h^{S,T,\kappa}) 
    \leq N(\lambda,\tilde M_h^{S,T,\kappa}).
  \end{equation*}
\end{proposition}

%%%%%%%%%%%%%%%%%%%%%%%%%%%%%%%%%%%%%%%%%%%%%%%%%%%%%%%%%%%%%%%%%%%%%%%%%%%%%%%

\subsection{Estimates near the boundary}\label{sec:bdryest2}

Similarly as in Subsection \ref{sec:bdryest} we will now approximate
the quadratic form $q_h$ locally near the boundary but, the magnetic
field now being constant, this can be done with a higher precision. In
particular we will see the curvature of the boundary appear. Again we
shall use the notation from Subsection \ref{sec:bdrycoord} and, in
contrast to Subsection \ref{sec:bdryest}, it will be important now to
keep $\tilde\kappa$ arbitrary.

\emph{We will assume in this subsection that $B\equiv 1$.} Then we can
choose the magnetic vector potential in the following way.

\begin{lemma}\label{gauge2}
  There exists a constant $C>0$ such that for all $S\in(0,L)$, $\tilde
  S\in[0,S]$ there exists a $\phi\in C^2([0,S]\times[0,t_0])$ such
  that
  \begin{equation*}
    \mathbf{\tilde A}(s,t)-\mathbf\nabla\phi(s,t)=
    (-t+\tilde\kappa t^2/2 + \beta(s,t),0)^T,
    \qquad (s,t)\in[0,S]\times[0,t_0],
  \end{equation*}
  where $\tilde\kappa:=\kappa(\tilde S)$ and for any $0<T\leq t_0$
  \begin{equation}\label{eq:gaugeremainder2}
    \sup_{(s,t)\in[0,S]\times[0,T]}|\beta(s,t)| \leq CS T^2.
  \end{equation}
\end{lemma}

Indeed, one can take $\phi(s,t):=\int_0^t \partial_s \tilde
A_2(s,t')\,dt' - \int_0^s \tilde A_1(s',0)\,ds'$ and recall that
$\partial_s \tilde A_2(s,t)- \partial_t \tilde A_1(s,t)=1-t\kappa(s)$.

Before stating the next result we recall the definition of $v$ from
\eqref{eq:uv} and of the quadratic form $m_h^{S,T,\kappa}$ from
Subsection \ref{sec:strip2}.

\begin{lemma}\label{bdrycoord2}
  Let $D>0$. Then there exists a constant $C>0$ such that for all
  $S\in(0,L)$, $\tilde S\in[0,S]$, $T\in(0,t_0]$ with
  $T\geq D\sqrt h$ and for all $u\in\mathcal{D}[q_h]$ such that the
  corresponding $v$ satisfies \eqref{eq:suppv} one has
  \begin{equation*}
    \begin{split}
      & \left| q_h[u]- m_h^{S,T,\tilde\kappa}[e^{-i\phi/h} v] \right|
      \\
      & \qquad \leq C \left( ST m_h^{S,T,\tilde\kappa}[e^{-i\phi/h} v]
      + (h^2T+ ST^3) \|e^{-i\phi/h}v\|_{\tilde\kappa}^2
      \right).
    \end{split}
  \end{equation*} 
  Here $\tilde\kappa:=\kappa(\tilde S)$ and $\phi$ is the function
  from Lemma \ref{gauge2}. Moreover, $\|\cdot\|_{\tilde\kappa}$ denotes
  the norm in $L_2((0,S)\times(0,T), a_{\tilde\kappa}dsdt)$.
\end{lemma}

\begin{proof}
  The proof is rather similar to that of Lemma \ref{bdrycoord}, so
  we will only sketch the major steps. Writing $w:=e^{-i\phi/h}v$ and
  taking \eqref{eq:newcoord} into account we decompose
  \begin{equation*}
    q_h[u]- m_h^{S,T,\tilde\kappa}[w] = I_1+I_2+I_3
  \end{equation*}
  where we define
  \begin{equation*}
    \begin{split}
      I_1 & :=  \int_0^\mathcal L \int_0^{t_0}
      \left(a^{-2}|(hD_s-\tilde{A_1})v|^2+|(hD_t-\tilde{A_2})v|^2\right)
      \,a_{\tilde\kappa}\,dsdt \\
      & \qquad - m_h^{S,T,\tilde\kappa}[w],\\
      I_2 & := h^2 \int_0^\mathcal L \int_0^{t_0} 
      W_{\tilde\kappa}|v|^2\,a_{\tilde\kappa}\,dsdt, \\
      I_3 & := -\frac{h^2}{2}\int_0^\mathcal L 
      (\kappa-\tilde\kappa) |v(.,0)|^2\,ds.
    \end{split}
  \end{equation*}
  To treat the terms $I_2$ and $I_3$ we use that
  $|\kappa-\tilde\kappa|\leq c_1 S$ on the support of $w$. This leads
  to the estimates
  \begin{equation*}
    \begin{split}
      |I_2| & \leq c_2 h^2(S+T) \|w\|_{\tilde\kappa}^2,\\
      |I_3| & \leq \epsilon \|hD_t w\|_{\tilde\kappa}^2 + 
      c_3\epsilon^{-1}h^2 S^2 \|w\|_{\tilde\kappa}^2.
    \end{split}
  \end{equation*}
  for any $\epsilon>0$. To take care of $I_1$ we note that by Lemma
  ~\ref{gauge2} one has
  \begin{equation*}
    \begin{split}
      I_1 & = \int_0^\mathcal L \int_0^{t_0}
      \left(a^{-2}|(hD_s+t-\tilde\kappa t^2/2-\beta)w|^2 \right. \\
      & \qquad\qquad\qquad \left.
      -a_{\tilde\kappa}^{-2}|(hD_s+t-\tilde\kappa t^2/2)w|^2\right)
      \,a_{\tilde\kappa}\,dsdt.
    \end{split}
  \end{equation*}
  Using the estimate on $\beta$ from \eqref{eq:gaugeremainder2} and
  that $|a^{-2}a_{\tilde\kappa}^2 - 1|\leq c_4 ST$ on the support of
  $w$ we easily find, for all $\epsilon>0$,
  \begin{equation*}
    \begin{split}
      |I_1| 
      & \leq (\epsilon + c_5 ST)
      \int_0^\mathcal L \int_0^{t_0}
      a_{\tilde\kappa}^{-2}|(hD_s+t-\tilde\kappa t^2/2)w|^2
      \,a_{\tilde\kappa}\,dsdt \\
      & \qquad
      + c_5(1+\epsilon^{-1})S^2T^4 \|w\|_{\tilde\kappa}^2.
    \end{split}
  \end{equation*}
  The assertion then follows by choosing $\epsilon=ST$ and recalling
  $T\geq D\sqrt h$.
\end{proof}

%%%%%%%%%%%%%%%%%%%%%%%%%%%%%%%%%%%%%%%%%%%%%%%%%%%%%%%%%%%%%%%%%%%%%%%%%%%%%%%

\subsection{Bracketing}\label{sec:bracketing2}

Now we estimate $N(h\Theta_0+h^{3/2}\mathcal C_1\kappa_0,P_h)$ by the
spectral counting functions of the operators with constant curvature
from Subsection ~\ref{sec:halfcylinder2}. Again we assume that
$B\equiv 1$.

For $N\in\N$ we define $S$ and $s_n$ as in \eqref{eq:s}. In contrast
to Subsection ~\ref{sec:bracketing} we will not yet specify the value
of $N$ but postpone this to the next subsection.

\begin{proposition}\label{bracketing2}
  Let $\kappa_0\in\R$. Under the above hypotheses there exists a
  constant $C>0$ such that for any $0<h\leq C^{-1}$,
  $\delta\in(0,S/2]$, $T$ with $Ch^{1/2}|\log h|\leq T \leq
  C^{-1}h^{1/4}$ and $\tilde S_n\in [s_{n-1},s_n]$, $n=1,\ldots,N$,
  one has
  \begin{equation*}\label{eq:bracketing2}
    \begin{split}
      & \frac12 \sum_{n=1}^N N(h\Theta_0+h^{3/2}\mathcal C_1\kappa_0-
      C(ST^3+h^2\delta^{-2}), 
      \tilde M_h^{2(S+\delta),T,\tilde\kappa_n}) \\
      & \qquad \leq N(h\Theta_0+h^{3/2}\mathcal C_1\kappa_0,P_h) \\
      & \qquad \leq \sum_{n=1}^N N(h\Theta_0+h^{3/2}\mathcal
      C_1\kappa_0+ 
      C(ST^3+h^2\delta^{-2}), 
      \tilde M_h^{S+2\delta,T,\tilde\kappa_n})
    \end{split}
  \end{equation*}
  where $\tilde\kappa_n:=\kappa(\tilde S_n)$.
\end{proposition}

The proof is similar to that of Proposition \ref{bracketing}, where
however Proposition ~\ref{strip} and Lemma \ref{gauge} are to be
replaced by Proposition \ref{strip2} and Lemma ~\ref{gauge2}
respectively. We omit the details.

%%%%%%%%%%%%%%%%%%%%%%%%%%%%%%%%%%%%%%%%%%%%%%%%%%%%%%%%%%%%%%%%%%%%%%%%%%%%%%%

\subsection{Proof of Theorem \ref{main2}}

Replacing $h$ by $h/B$ we may assume that $B\equiv 1$. We will show that
\begin{equation}\label{eq:main2remainder}
  \begin{split}
    h^{1/4} N(h\Theta_0+h^{3/2}\mathcal C_1\kappa_0,P_h) 
    & = \frac 1{\pi\sqrt{3|\xi_0|}} \int_0^\mathcal L
    (\kappa(s)+\kappa_0)_+^{1/2}\,ds \\
    & \qquad +\mathcal O_\epsilon(h^{1/16-\epsilon})
  \end{split}
\end{equation}
for any $\epsilon>0$. As in Subsection \ref{sec:proofmain} we give the
proof of the lower bound only.

Let $N$, $S$ be as in Subsection \ref{sec:bracketing2} and let $\tilde
S_n\in [s_{n-1},s_n]$ be arbitrary with $\tilde\kappa_n:=\kappa(\tilde
S_n)$. From Proposition \ref{halfcylinder2} and Proposition
\ref{bracketing2} we get for all $\delta\in(0,S/2]$, $Ch^{1/2}|\log
h|\leq T \leq C^{-1}h^{1/4}$ (with $C$ as in Proposition
\ref{bracketing2})
\begin{equation}\label{eq:proofmain2est1}
  \begin{split}
    & h^{1/4} N(h\Theta_0+h^{3/2}\mathcal C_1\kappa_0,P_h) \\
    & \qquad \geq (\pi\sqrt{3|\xi_0|})^{-1} (S-\delta)
    \sum_{n=1}^N
    (\tilde\kappa_n+\kappa_0-c_1(h^{-3/2}ST^3+h^{1/2}\delta^{-2}))_+^{1/2}
    \\
    & \qquad \qquad -c_2 N h^{1/4}.
  \end{split}
\end{equation}
Now we use the estimate
\begin{equation*}
  \begin{split}
    & (\tilde\kappa_n+\kappa_0-
    c_1(h^{-3/2}ST^3+h^{1/2}\delta^{-2}))_+^{1/2} \\
    & \qquad \geq (\tilde\kappa_n+\kappa_0)_+^{1/2}
    -c_1^{1/2}(h^{-3/4}S^{1/2}T^{3/2}+h^{1/4}\delta^{-1}).
  \end{split}
\end{equation*}
If we assume that $\delta\geq h^{1/4}$ we easily deduce from
\eqref{eq:proofmain2est1} that
\begin{equation}\label{eq:proofmain2est2}
  \begin{split}
    h^{1/4} N(h\Theta_0+h^{3/2}\mathcal C_1\kappa_0,P_h)
    & \geq (\pi\sqrt{3|\xi_0|})^{-1} S
    \sum_{n=1}^N (\tilde\kappa_n+\kappa_0)_+^{1/2}
    \\
    & \qquad -c_3
    (h^{-3/4}S^{1/2}T^{3/2}+h^{1/4}\delta^{-1}+\delta S^{-1}).
  \end{split}
\end{equation}
Now choose $T= h^{1/2-\rho}$ with $0<\rho<1/4$. A calculation shows
that the second term on the RHS of \eqref{eq:proofmain2est2} is
minimal for the choice
\begin{equation*}
  S= h^{1/8+3\rho/2}, \qquad \delta = h^{3/16+3\rho/4}
\end{equation*}
and given by $c_4 h^{1/16-3\rho/4}$. The first term on the RHS
of \eqref{eq:proofmain2est2} is a Riemannian sum. Recalling that the
$\tilde S_n$ were arbitrary, we finally arrive at the lower bound of
\eqref{eq:main2remainder}. The upper bound can be established
similarly, which concludes the proof of Theorem \ref{main2}.

%%%%%%%%%%%%%%%%%%%%%%%%%%%%%%%%%%%%%%%%%%%%%%%%%%%%%%%%%%%%%%%%%%%%%%%%%%%%%%%
%%%%%%%%%%%%%%%%%%%%%%%%%%%%%%%%%%%%%%%%%%%%%%%%%%%%%%%%%%%%%%%%%%%%%%%%%%%%%%%
%%%%%%%%%%%%%%%%%%%%%%%%%%%%%%%%%%%%%%%%%%%%%%%%%%%%%%%%%%%%%%%%%%%%%%%%%%%%%%%

\bibliographystyle{amsalpha}

\end{document}